\documentclass[twocolumn,amsmath,amssymb]{revtex4}

\usepackage{graphicx}
\usepackage{dcolumn}
\usepackage{bm}

\def\boldgrad{\bm\nabla}
\def\grad{\nabla}

\begin{document}

\title{Vortex dynamics in the nonlinear Schr\"odinger equation}
\author{Michael J. Quist}
\email{mjq1@cornell.edu}
\affiliation{Department of Physics, Cornell University, Ithaca, New York 14850}

\date{\today}

\begin{abstract}
The dynamics of a two-dimensional vortex are analyzed within the
framework of the nonlinear Schr\"odinger equation.  Both a bare vortex
and a vortex with an external mass trapped in a finite-sized core are
considered.  The bare vortex motion is found to be damped at all
frequencies, while the finite core has a single resonant frequency.
The force exerted by the fluid on the finite core can be expressed as a
sum of dissipative and Magnus forces for sufficiently low frequencies,
even when the core is small.
\end{abstract}

\maketitle

\section{Introduction}

Vortices represent an important class of excitations in many-body 
systems.  They are associated with quantized circulation and 
dissipative flow in superfluids~\cite{1}, flux penetration and the 
breakdown of superconductivity in type-II superconductors~\cite{2}, and 
the Kosterlitz-Thouless phase transition in two-dimensional 
systems~\cite{KT}.  The study of vortices in classical fluids is also 
well-developed, and relevant material can be found in any good
hydrodynamics textbook.  Finally, the interested reader may wish
to consult Ref.~\cite{vorticesnonlinear}, a recent monograph covering
vortices, vortex lines, and vortex dynamics in all manner of nonlinear
media.

The aim of this paper is to address two questions 
pertaining to vortex dynamics in a two-dimensional boson fluid.  
First, we would like to determine whether there is an undamped or 
weakly damped mode with a moving vortex core; here the core velocity 
is measured relative to the fluid at infinity.  Second, we want to 
understand the dynamics of the core when an external mass is trapped 
in it.  A commonly-used phenomenological model expresses the force on 
a moving vortex core as the sum of a damping force acting parallel to 
the core velocity (again, relative to distant fluid) and a Magnus 
force acting perpendicular to it~\cite{magnusmodel}.  This type of model 
can describe classical vortex dynamics; it is also experimentally 
known to be valid in superfluid ${}^{4}$He, at least when the core is 
a macroscopic object~\cite{expevidence}.  We would like to know whether 
such a model also describes the dynamics at shorter length scales, on 
the order of the size of a bare vortex core.  We work within the 
nonlinear Schr\"odinger approximation, which is applicable in the 
limit of a dilute, weakly interacting Bose gas.  A recent publication 
by Demircan, Ao, and Niu addressed similar questions within this 
framework~\cite{demircan96}; we found that this work contained several 
mathematical errors invalidating its main results.

The outline of the paper is as follows.  We first present the basic 
vortex model and derive equations of motion for the core and the 
phonon modes to which it couples.  We then show how these equations 
can be modified to include a massive object in the core.  Next we 
discuss the qualitative features of the normal modes of the system and 
use numerical methods to find them.  Finally we present our results 
and conclusions.

\section{Vortex model}

We start with the following Lagrangian~\cite{negeleorland},
which describes a 
two-dimensional system of bosons interacting via a
delta-function pseudopotential:
\begin{equation}
  L = \int \! d^{2}r \left\{ i\hbar\Psi^{*}\frac{\partial\Psi}{\partial t} - 
  \frac{\hbar^{2}}{2m}{|\boldgrad\Psi|}^{2} + 
   \mu{|\Psi|}^{2} - \frac{\lambda}{2}{|\Psi|}^{4} \right\}.
\end{equation}
Here $m$ is the boson mass, $\mu$ the chemical potential, and 
$\lambda$ the interaction strength.  This can be put in dimensionless 
form by a rescaling of variables.  We let $\Psi\to 
({\mu}/{\lambda})^{1/2}\Psi$, ${\bf r}\to 
({\hbar^{2}}/{m\mu})^{1/2}{\bf r}$, and $t\to ({\hbar}/{\mu})t$ to 
obtain
\begin{equation}  
  L = \frac{\hbar^{2}\mu}{m\lambda}
   \int \! d^{2}r \left\{ i\Psi^{*}\frac{\partial\Psi}{\partial t} - 
     \frac{1}{2}{|\boldgrad\Psi|}^{2} + 
       {|\Psi|}^{2} - \frac{1}{2}{|\Psi|}^{4} \right\}.
\end{equation}
The prefactor is irrelevant for the present (classical) analysis, 
since the classical equations of motion are invariant under a rescaling
of $L$.  However, note that the action $\int \!  L \, dt$ 
acquires a prefactor of ${\hbar^{3}}/{m\lambda}$, which is much larger 
than $\hbar$ provided that ${\lambda}\ll{\hbar^{2}}/{m}$; this defines 
the weakly interacting limit.  In this limit the correspondence 
principle applies, at least naively, and quantum fluctuations around 
the classical behavior are expected to be small.  The Euler-Lagrange 
equation for the complex field $\Psi ({\bf r},\,t)$ is
\begin{equation}
 i\frac{\partial\Psi}{\partial t} = -\frac{1}{2}\grad^{2}\Psi + \left(|\Psi|^{2} - 1\right)\Psi.
\label{NLSE}
\end{equation}
\pagestyle{headings}%
This is the well-known nonlinear Schr\"odinger equation, derived for
the imperfect Bose gas by Gross and Pitaevskii~\cite{gross,pitaevskii}.
It has since been applied to vortex lines in superfluids by a number
of authors (see, e.g., Refs.~\cite{fetter,fetter2,pathria66}).
It can be given a hydrodynamic interpretation by making the Madelung 
transformation $\Psi=\sqrt{\rho}e^{i\phi}$, where $\rho$ and $\phi$ 
are the fluid density and velocity potential (${\bf 
v}=\boldgrad\phi$) respectively.  The equations of motion for 
these variables are
\begin{subequations}
\begin{eqnarray}
\frac{\partial\rho}{\partial t} &=& -\boldgrad\cdot(\rho{\bf v}), \\
\frac{d}{dt}(\rho{\bf v}) &=& \boldgrad\cdot{\bm\sigma},
\end{eqnarray}
\end{subequations}
where $d/dt\equiv\partial/\partial t+{\bf v}\cdot\boldgrad$ is the 
convective derivative and the components of the stress tensor
${\bm\sigma}$ are
\begin{equation}
\label{stresstensor}
\sigma_{ij} = -\frac{1}{2}(\rho^{2}-1)\delta_{ij} + 
\frac{1}{4}\left(\partial_{i}\partial_{j}\rho - 
\frac{1}{\rho}(\partial_{i}\rho)(\partial_{j}\rho)\right).
\end{equation}
Except for the derivative terms in the stress tensor, these are the 
equations of motion for an ideal classical fluid with pressure 
$p(\rho)=\frac{1}{2}(\rho^{2}-1)$.

Equation (\ref{NLSE}) has time-independent vortex solutions of the 
form $f(r)e^{in\theta}$ for any integer $n$.  We consider only $n=1$, 
in which case the solution is
\begin{equation}
\Psi_{0}({\bf r}) = f(r)e^{i\theta},
\end{equation}
where $f(r)$ satisfies
\begin{equation}
f''(r) + \frac{1}{r}f'(r) + 
    \left(2 - 2f(r)^{2} - \frac{1}{r^{2}}\right)f(r) = 0.
\label{fdef}
\end{equation}
The asymptotic behaviors of the solution for large and small $r$ can 
be expressed as power series:
\begin{equation}
\label{eqn:fasymptotic}
f(r) \sim
\begin{cases}
         A\,r - \frac{A}{4}\,r^{3} + 
         \left(\frac{A}{48} + \frac{A^{3}}{12}\right) \, r^{5} 
         +\cdots & {\rm as\;\;} r \to 0,
\cr
         1 - \frac{1}{4\,r^{2}} - \frac{9}{32\,r^{4}} - \cdots
         & {\rm as\;\;} r\to\infty.
\cr
\end{cases}
\end{equation}
The value of $A\equiv f'(0)$ is not determined by the asymptotics and 
must be found numerically.  We used a shooting 
method and found $A\simeq 0.825$, which does not agree with the value 
of $\sqrt{2}$ given in Ref.~\cite{demircan96}, but does agree
with the numerical work of Kawatra and Pathria~\cite{pathria66}.

In order to investigate vortex motion, we consider perturbations
around a single-vortex solution with a moving core.  The new dynamical
variables are defined by
\begin{equation}
\label{eqn:newvars}
  \Psi({\bf r}, t) = \Psi_{0}\Bigl({\bf r} - {\bf r}_{0}(t)\Bigr) +
  \delta\Psi\Bigl({\bf r} - {\bf r}_{0}(t),\,t\Bigr),
\end{equation}
where ${\bf r}_{0}(t)$ is the location of the vortex core.  Changing
variables in Eq.~(\ref{NLSE})
and linearizing in $\delta\Psi$ and $d{\bf r}_{0}/dt$
gives
\begin{eqnarray}
\label{evolutioneqn}
i\frac{\partial\delta\Psi}{\partial t} &=& i\frac{d{\bf r}_{0}}{dt}\cdot\boldgrad\Psi_{0} -
      \frac{1}{2}\grad^{2}\delta\Psi 
       + \left(2|\Psi_{0}|^{2}-1\right)\delta\Psi \nonumber\\ &&+ 
      \left(\Psi_{0}\right)^{2}\delta\Psi^{*}.
\end{eqnarray}
We express $\delta\Psi$ as a sum over cylindrical harmonics: 
$\delta\Psi({\bf r}) = \sum_{m}\delta\Psi_{m}(r)e^{im\theta}$.  It is 
also convenient to represent ${\bf r}_{0}$ by a complex number 
$r_{+}\equiv{\bf r}_{0}\cdot(\hat{\bf x}+i\hat{\bf y})$.  In Eq.~(\ref{evolutioneqn}),
only the $m=0$ and $m=2$ modes couple to the dynamics of 
${\bf r}_{0}$.  The other modes can be set to zero, and the evolution of
the relevant modes is given by
\begin{subequations}
\label{modes02eqns}
\begin{eqnarray}
\label{mode0eqn}
i\frac{{\partial\delta\Psi}_{0}}{\partial t} &=& -\frac{1}{2}\delta\Psi_{0}'' - 
\frac{1}{2r}\delta\Psi_{0}'
\nonumber\\
&&\;\;\; + \left(2f^{2}-1\right)\delta\Psi_{0} + f^{2}\delta\Psi_{2}^{*}
\nonumber\\
&&\;\;\; + \frac{1}{2}i\frac{d{r}_{+}}{dt}\left(f'+\frac{f}{r}\right),  \\
\label{mode2eqn}
-i\frac{\partial\delta\Psi_{2}^{*}}{\partial t} &=& -\frac{1}{2}(\delta\Psi_{2}^{*})'' - 
\frac{1}{2r}(\delta\Psi_{2}^{*})'
\nonumber\\
&&\;\;\; + \left(2f^{2}-1+\frac{2}{r^{2}}\right)\delta\Psi_{2}^{*} + f^{2}\delta\Psi_{0}
\nonumber\\
&&\;\;\; - \frac{1}{2}i\frac{d{r}_{+}}{dt}\left(f'-\frac{f}{r}\right),
\end{eqnarray}
\end{subequations}
where the primes represent differentiation with respect to $r$.  We 
impose the constraint $\delta\Psi({\bf 0},\,t)=0$, so that ${\bf 
r}_{0}$ marks the true location of the vortex core.  These equations 
then determine the motion of both the fluid and the core.  In 
particular, since the $\delta\Psi_{m}$ and their time derivatives 
vanish as $r\to 0$, Eq.~(\ref{mode0eqn}) implies that
\begin{equation}
\label{rdoteqn}
\frac{d{r}_{+}}{dt} = -\frac{i\delta\Psi_{0}''(0)}{f'(0)}.
\end{equation}

\section{Finite core}

\begin{figure}
\includegraphics[width=3.25 in]{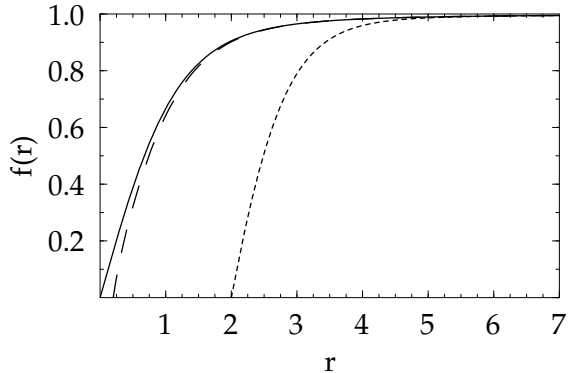}
\caption{\label{fig:fgraph}Square root of the fluid density for the bare core (solid
curve), small core (dashed curve), and large core (dotted curve).}
\end{figure}

We also would like to introduce an additional mass $M$, coupled to an 
arbitrary external force ${\bf F}(t)$ and confined to the center of 
the vortex.  The question arises as to how this mass (which might 
physically represent a foreign particle trapped in the core) should 
interact with the fluid.  We considered using a point mass constrained 
to lie at the vortex core, but found this model to be internally 
inconsistent for the following reason.  If the perturbation 
$\delta\Psi$ is bounded, as the validity of our linearized approach 
requires, the fluid exerts no force on a point mass.  (This is 
justified in the next paragraph.)  The equation of motion for a point 
mass is therefore $M\ddot{\bf r}_{0}={\bf F}$\null.  On the other 
hand, for the mass to remain at the vortex core, Eq.~(\ref{rdoteqn}) needs to be 
satisfied as well.  The two equations for ${\bf r}_{0}$ cannot be 
satisfied simultaneously for arbitrary ${\bf F}$, so the model is 
ill-defined.  Instead, we will treat the core as a hard disk of finite 
radius $a$.  This is a simple approximation to a more realistic 
interaction between the fluid and a trapped particle, such as a 
Lennard-Jones potential.  The fluid density must vanish inside the 
core, so the stationary vortex solutions now satisfy $f(a)=0$, and 
$\delta\Psi({\bf r},\,t)=0$ for $r=a$.  Figure~\ref{fig:fgraph} shows
$f(r)$ for the three cases we considered: the bare core, a small
core ($a=0.2$), and a large core ($a=2.0$).  The time dependence
of small perturbations in 
the fluid is still given by Eqs.~(\ref{modes02eqns}).

For momentum to be locally conserved, the force on the core due to the 
fluid must equal the momentum crossing the boundary line $r=a$ per 
unit time.  This condition can be expressed in terms of the stress tensor:
\begin{equation}
\label{coreforceeqn}
\frac{d{\bf F}^{\rm fluid}}{ds} = {\bm\sigma}\cdot\hat{\bf n},
\end{equation}
where $\hat{\bf n}$ is the boundary's outward unit normal and $ds$ is 
the line element along the boundary.  Integrating this expression 
around the circumference of the core and using the form of the stress 
tensor given in Eq.~(\ref{stresstensor}), the force exerted on the mass by the fluid is 
found to be
\begin{eqnarray}
F^{\rm fluid}_{+}&\equiv&{\bf F}^{\rm fluid}\cdot
(\hat{\bf x}+i\hat{\bf y}) \nonumber\\ &=&
                    \pi a f'(a)\Bigl(
                    \delta\Psi_{0}'(a)+(\delta\Psi_{2}^{*})'(a)\Bigr).
\end{eqnarray}
Note that as $a\to 0$, this force vanishes unless 
$\delta\Psi_{0}+\delta\Psi_{2}^{*}$ diverges at least logarithmically, 
as asserted above.  Here we mention that Ref.~\cite{demircan96} 
draws its conclusions from the point-mass model, which we have argued 
to be inconsistent.  Their presentation of this model differs from ours in 
that they introduce a mass term directly into the Lagrangian.  
This hides the inconsistency but does not remove it: it leads to a 
seemingly nontrivial equation of motion for ${\bf r}_{0}$ which is 
actually equivalent to $M\ddot{\bf r}_{0}={\bf 0}$.  A subsequent 
error (in joining inner and outer asymptotic solutions) disguises this 
equivalence, enabling the derivation of quantitative results that are
essentially unsupported.

The equation of motion for the core mass is now
\begin{equation}
\label{coremotioneqn}
M{\ddot r}_{+} = F^{\rm fluid}_{+} + F_{+},
\end{equation}
where the external force has been written as a complex number 
$F_{+}\equiv{\bf F}\cdot(\hat{\bf x}+i\hat{\bf y})$.  Since the 
equations of motion (Eqs.~(\ref{modes02eqns}) and (\ref{coremotioneqn}))
are linear in $F_{+}$, $r_{+}$, $\delta\Psi_{0}$, and $\delta\Psi_{2}^{*}$,
we will seek solutions in which all of these variables are proportional to 
$e^{i\omega t}$.  The general solution can be then expressed as a sum 
over normal modes in the usual way.

\section{Normal modes}
\label{sec:normalmodes}

For a finite core, the mode equations for frequency $\omega$ are
\begin{widetext}
\begin{subequations}
\label{eqn:omegamodes}
\begin{eqnarray}
-M\omega^{2}r_{+} - \pi a f'(a)\Bigl(
      \delta\Psi_{0}'(a)+(\delta\Psi_{2}^{*})'(a)\Bigr) &=& F_{+}, \\
\delta\Psi_{0}'' + \frac{1}{r}\delta\Psi_{0}' + 
\left(2-4f^{2}-2\omega\right)\delta\Psi_{0} - 
2f^{2}\delta\Psi_{2}^{*} &=&
-\omega r_{+}\left(f'+\frac{f}{r}\right),   \nonumber\\ \\
(\delta\Psi_{2}^{*})'' + \frac{1}{r}(\delta\Psi_{2}^{*})' + 
\left(2-4f^{2}-\frac{4}{r^{2}}+2\omega\right)\delta\Psi_{2}^{*} - 
2f^{2}\delta\Psi_{0} &=& \omega r_{+}\left(f'-\frac{f}{r}\right); \nonumber\\
\end{eqnarray}
\end{subequations}
\end{widetext}
the bare core will be considered as a special case with $a=M=F_{+}=0$.
Equations (\ref{eqn:omegamodes}b-c) are coupled inhomogeneous linear ODEs 
for $\delta\Psi_{0}(r)$ and $\delta\Psi_{2}^{*}(r)$ which must be 
solved subject to homogeneous boundary conditions: the 
$\delta\Psi_{m}$ must remain finite as $r\to\infty$, and must vanish 
as $r\to a$.  The most general solution will have the form of a 
particular solution to the inhomogeneous equations plus a linear 
combination of solutions to the corresponding homogeneous equations 
(obtained by setting $r_{+}=0$).

We now consider the solutions to the homogeneous equations.  Given the 
order of the equations, there must be four such solutions.  As $r\to 
0$ their asymptotic behaviors are as follows: (a) $\delta\Psi_{0}\sim 
r^{6}$ and $\delta\Psi_{2}^{*}\sim const.\times r^{2}$; (b) 
$\delta\Psi_{0}\sim 1$ and $\delta\Psi_{2}^{*}\sim const.\times 
r^{4}$; (c) $\delta\Psi_{0}\sim\ln r$ and $\delta\Psi_{2}^{*}\sim 
const.\times r^{4}\ln r$; and (d) $\delta\Psi_{0}\sim r^{2}$ and 
$\delta\Psi_{2}^{*}\sim const.\times 1/r^{2}$.  In the opposite limit, 
as $r\to\infty$, two solutions are oscillatory, representing incoming 
and outgoing waves, one solution decays exponentially, and one 
solution grows exponentially.  (The asymptotic solutions were found by 
Fetter~\cite{fetter}.) All other points ($r\ne 0$ or $\infty$) are 
regular points of Eqs.~(\ref{eqn:omegamodes}b-c), so the $\delta\Psi_{m}$ and their 
derivatives can be specified freely.

In the case of the bare core, only inner solution (a) conforms to the 
boundary condition at the origin.  If this function is followed to 
infinity, it will decompose into a linear combination of the four 
outer solutions, and in general (for arbitrary $\omega$) the 
coefficient of the exponentially growing solution will not vanish.  In 
other words, if the inner boundary condition is satisfied then the 
outer one will not be.  Therefore, with a bare core there is no 
nonzero solution to the homogeneous equations, and the solution to the 
inhomogeneous equations will be unique and proportional to $r_{+}$.

In the case of a finite core, two linearly independent solutions
satisfy the boundary condition at $r=a$, since $\delta\Psi_{0}'(a)$
and $(\delta\Psi_{2}^{*})'(a)$ are independent free parameters.
Both have some coefficient of overlap with the growing 
solution, so a linear combination can be formed for
which the overlap is zero and both boundary conditions are 
satisfied.  The general solution to the inhomogeneous equations is in 
this case the sum of a term proportional to $r_{+}$ and a term 
proportional to another free parameter.  Physically we expect to find 
a unique solution where the mass is driven solely by the external force and
the fluid contains only outgoing waves.  This solution should be 
proportional to $r_{+}$.  The free parameter associated with the
homogeneous solution can be taken to be the amplitude of incoming waves.
Once the solutions are found, the 
force exerted on the moving mass by the fluid is given by
Eq.~(\ref{coreforceeqn}).

At this point we resort to numerical techniques to determine the
solutions for the two cases.

\section{Numerical methods and results}

No single numerical technique proved capable of solving Eqs.~(\ref{eqn:omegamodes}b-c) 
over the entire range $a<r<\infty$.  Instead, a different method was 
used for each of three regions: $a<r<a+4$, $a+4<r\lesssim 
2\pi/\omega$, and $2\pi/\omega \lesssim r<\infty$.  In each region we 
found a single solution to the inhomogeneous equations and all four 
linearly independent solutions to the homogeneous equations.  We then 
chose the appropriate coefficients for the homogeneous terms so that 
$\delta\Psi_{0}$ and $\delta\Psi_{2}^{*}$ satisfied the boundary 
conditions and joined smoothly at the boundaries between regions.

For the innermost region we used direct numerical integration, starting at 
$r=a$, to determine the homogeneous solutions.  Because of the 
exponentially growing solution, the integration is unstable and loses 
precision with increasing $r$.  We cut off the integration at $r=a+4$; 
the exact placement was somewhat arbitrary, but it cannot be 
much further from the core.  For the outermost region we used an 
asymptotic analysis.  All the outer solutions to the homogeneous 
equations have asymptotic expansions of the form
\begin{equation}
\label{eqn:outerasymptotic}
\delta\Psi_{0}\sim \frac{e^{kr}}{\sqrt r}\sum_{n=0}^{\infty}a_{n}(k)r^{-n},
\;
\delta\Psi_{2}^{*}\sim \frac{e^{kr}}{\sqrt r}\sum_{n=0}^{\infty}b_{n}(k)r^{-n},
\end{equation}
where $k$ takes on one of the four values $\pm\sqrt{2 \pm 
2\sqrt{1+\omega^{2}}}$.  Using the full outer asymptotic series for 
$f(r)$, the first few terms of which are shown in Eq.~(\ref{eqn:fasymptotic}), the 
coefficients $a_{n}(k)$ and $b_{n}(k)$ can be found for arbitrarily 
large $n$.  However, we found $a_{1}/a_{0}$ and $b_{1}/b_{0}$ to be of order 
$1/\omega$ for small $\omega$, indicating that the $n=0$ terms in the 
series dominate for $r\gg 1/\omega$.  We therefore kept only the $n=0$ 
terms and cut off the outermost region at $r\simeq 2\pi/\omega$.  For 
the middle region we put Eqs.~(\ref{eqn:omegamodes}b-c) on a grid, discretizing the 
derivatives in the most straightforward way.  For small values of 
$\omega$ this middle region becomes quite large, so the average grid 
spacing needs to be reasonably large as well, to limit the number of 
grid points.  On the other hand, the solutions still vary rapidly near 
the boundary at $r=a+4$, so the spacing must be much smaller there.  A 
nonuniform grid was used to address both these problems.

Finally, note that an exact solution to the inhomogeneous equations, 
for all three regions,
is
\begin{equation}
\label{eqn:exactsoln}
\delta\Psi_{0}=\frac{1}{2}r_{+}\left(f'+\frac{f}{r}\right), \;\;
\delta\Psi_{2}^{*}=\frac{1}{2}r_{+}\left(f'-\frac{f}{r}\right).
\end{equation}
This solution has a straightforward physical interpretation: it 
describes a stationary
vortex centered at the origin.  To see this, we can refer back to 
Eq.~(\ref{eqn:newvars}), the defining equation for $\delta\Psi$.  A stationary vortex
is described by $\Psi({\bf r},\,t) = \Psi_{0}({\bf r})$, so
\begin{eqnarray}
\delta\Psi\Bigl({\bf r}-{\bf r}_{0}(t),\,t\Bigr) &=& \Psi_{0}({\bf r}) - 
\Psi_{0}\Bigl({\bf r}-{\bf r}_{0}(t)\Bigr) \nonumber\\
          &=& {\bf r}_{0}(t)\cdot\boldgrad\Psi_{0}({\bf r})
\end{eqnarray}
to first order in ${\bf r}_{0}$, and the angular components
of this function are indeed given by Eq.~(\ref{eqn:exactsoln}).

We found that our procedure gave good results for a wide range of 
frequencies.  The results were extremely insensitive to adjustment of 
the region boundaries, increased precision in the numerical 
integration, and the inclusion of more terms in the asymptotic 
expansions.  Changing the number and spacing of grid points in the 
middle region, however, did cause small variations in the final 
results.  In particular, for $10^{-3}\lesssim |\omega|\lesssim 5$ we 
estimate our errors to be on the order of a few percent.

\begin{figure}[t]
\includegraphics[width=3.25 in]{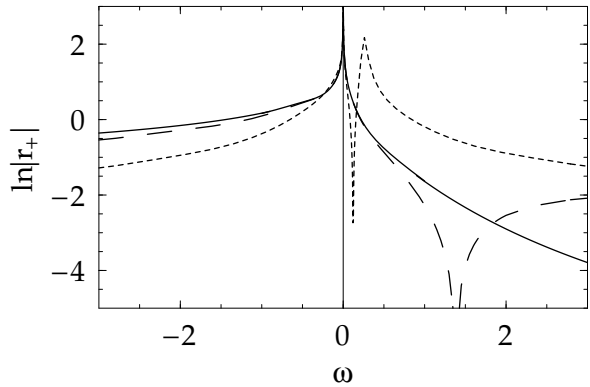}
\caption{\label{fig:fig2}Amplitude of circular motion in response to incoming waves
of unit power and frequency $\omega$.  Results are for the bare core 
(solid curve), small core (dashed curve), and large core (dotted 
curve).  ${\rm Ln}|r_{+}|$ is shown.}
\end{figure}

First we analyzed elastic scattering.  When incoming waves of unit 
power are scattered by a free core ($F_{+}=0$), the core responds 
with circular motion of a particular amplitude.  Figure~\ref{fig:fig2}
shows the amplitude of response as a function of frequency for the
bare, small, and large cores.  Note that for a finite core the response depends on 
the mass; we show the results for $M=0$ only, but using $M>0$ causes
no qualitative changes.  A divergent response is seen 
only as $\omega\to 0$.  The response functions for the small and large 
cores each have a single zero.  These correspond to resonances in 
which the driven motion (discussed below) is undamped and the core 
motion is decoupled from the phonon field.  The response function for 
the bare core, in contrast, is everywhere nonzero; we conclude that 
there is no undamped or weakly damped finite-frequency mode.

\begin{figure}[t]
\includegraphics[width=3.25 in]{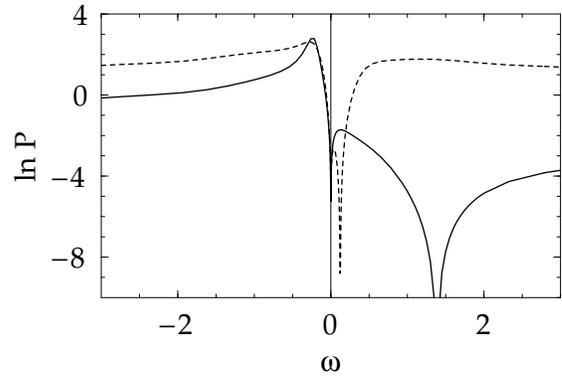}
\caption{\label{fig:fig3}Power dissipated by the small core (solid curve) and large 
core (dotted curve) in driven circular motion with unit speed and frequency 
$\omega$.  ${\rm Ln}P$ is shown.}
\end{figure}

For the small and large cores, we also analyzed driven motion.  When a 
finite core is driven by the external force with unit speed, power is 
radiated as outgoing waves.  The power is equal to that put into the 
system by the component of ${\bf F}$ parallel to the velocity: for 
$r_{+}=1/|\omega|$, corresponding to circular motion with unit speed, 
it is given by $P={\bf F}\cdot d{\bf r}_{0}/dt = \pm\,{\rm Im}F_{+}$, 
where the sign is $+/-$ for positive/negative frequencies.  This 
quantity is shown in Fig.~\ref{fig:fig3} as a function of frequency.  It is 
independent of $M$, and so provides a better measure of damping than 
the elastic scattering response.  For the small core, the radiated 
power vanishes at $\omega\simeq 1.5$.  At this frequency there are 
neither incoming nor outgoing waves; the perturbation is localized and 
dies off exponentially with increasing $r$.  There is a similar 
resonance for the large core at $\omega\simeq 0.13$.  For both cores, 
the motion is undamped as $\omega\to 0$.  Results for other core sizes 
(not shown) suggest that these features are general: there is always a 
single positive-frequency resonance, the frequency of which varies 
inversely with $a$, and the damping always approaches zero for low 
frequencies.

\begin{figure}[t]
\includegraphics[width=3.25 in]{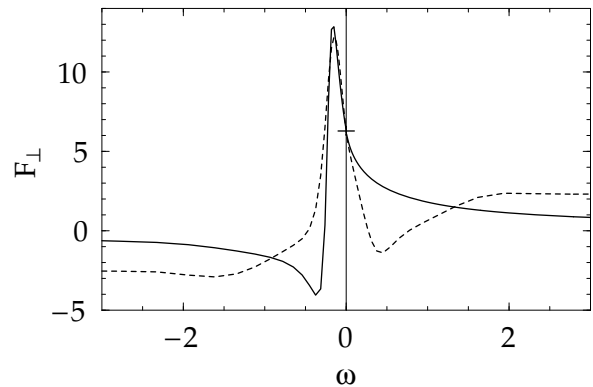}
\caption{\label{fig:fig4}Force in the $-d{\bf r}_{0} / dt \times \hat{\bf z}$ direction 
exerted by the fluid on the small core (solid curve) and large core 
(dotted curve) in driven circular motion with unit speed and frequency 
$\omega$.  The mark on the vertical axis corresponds to 
$F_{\perp}=2\pi$.}
\end{figure}

Figure~\ref{fig:fig4} shows the component of the force exerted by the fluid on the 
mass in the $-d{\bf r}_{0} / dt \times \hat{\bf z}$ direction.  If the 
only contributor were the Magnus force,
\begin{equation}
{\bf F}^{\rm Magnus} = -2\pi \frac{d{\bf r}_{0}}{dt} \times \hat{\bf z},
\end{equation}
this component would be identically $2\pi$, which is marked on the 
vertical axis.  Because we are driving the core with unit speed, the 
dependence on $\omega$ should vanish.  Instead we see that the Magnus 
effect dominates only in the limit as $\omega\to 0$; for 
$|\omega|\gtrsim 1/10$, it can no longer account for the perpendicular 
component of the force.

\section{Conclusions}

From our data we can answer both of our original questions.  First, 
there is no undamped mode associated with the motion of a bare vortex.  
Second, the Magnus effect plays an important role in the dynamics of a 
finite core for the frequency range $|\omega|\lesssim 1/10$, even when 
the core is small.  The Magnus force has been shown to be a general 
consequence of vortex motion in the adiabatic phase 
approximation~\cite{ao93}.  We point out that this approximation is not 
valid when applied to the nonlinear Schr\"odinger equation, at least 
for an infinite system.  To see why it must fail, again consider the 
solutions to Eqs.~(\ref{eqn:omegamodes}b-c) in the limit as $r\to\infty$.  The 
particular solution given by Eq.~(\ref{eqn:exactsoln}), as already stated, corresponds 
to a \emph{stationary} vortex.  The phase field of distant fluid does 
not follow the instantaneous location of the vortex, as it does within 
the adiabatic phase approximation, but rather maintains a fixed 
center.  The most general solution to the mode equations has the same 
property, as the solutions to the corresponding homogeneous equations 
are either oscillatory or exponentially decaying, as discussed in 
Section~\ref{sec:normalmodes}, and cannot cancel this power-law behavior.
The failure of 
the adiabatic phase approximation is also expected on physical 
grounds: since the low-frequency phonons travel with finite speed (the 
speed of sound $c=1$ in our units), the fluid cannot respond to 
low-frequency disturbances at a distance $d$ in a time shorter than 
$1/d$.  For this response time to be shorter than the time scale 
associated with the disturbance, we must have $d\ll 1/\omega$; at 
larger distances the phase of the fluid can no longer follow the core.  
Nevertheless we recover the Magnus effect for low frequencies, in 
agreement with results dependent on the adiabatic phase approximation, 
suggesting that the discrepancy at large distances is not crucial.  
For high frequencies, the Magnus force ceases to play a role.  Note 
that this transition occurs while the phonon wavelength is still very 
large compared to the size of the core: for $\omega=1/10$, the phonon 
wavelength is around 60.  For superfluid ${}^{4}$He, the transition 
frequency is 20-30 GHz, corresponding to a phonon wavelength of 100 
\AA.

\begin{acknowledgments}
This work was performed with the kind support of Cornell University.
\end{acknowledgments}


\end{document}